\documentclass{an}
\usepackage{graphicx}
\usepackage{times}
\usepackage{fancyhdr}
\def\apj{{ApJ }}
\def\apjs{{ ApJ~Suppl. }}

\def\aa{{A\&A }}

\def\an{{AN }}
\def\sp{{Solar Phys. }}
\begin{document}

\title{Differential rotation on the lower main sequence}

\author{M.~K\"uker and G.~R\"udiger}
\institute{Astrophysikalisches Institut Potsdam, An der Sternwarte 16, 14482 Potsdam, Germany}

\date{Received; accepted; published online}

\abstract{We compute the differential rotation of main sequence stars of the spectral types F, G, K, and M by solving the equation of motion and the equation of convective heat transport in a mean-field formulation. For each spectral type the rotation rate is varied to study the dependence of the surface shear on this parameter. The resulting rotation patterns are all solar-type. The horizontal shear turns out to depend strongly on the effective temperature and only weakly on the rotation rate. The meridional flow depends more strongly on the rotation rate and has different directions in the cases of very slow and very fast rotation, respectively.
}

\correspondence{mkueker@aip.de}

\maketitle

\section{Introduction}
The surface rotation of the Sun shows a shorter period at the equator than at the poles. Helioseismology has shown that this pattern persists throughout the convection zone and disappears at its bottom (Thompson et al.~2003). So far differential rotation can not be measured directly for other main-sequence stars. It has, however, been inferred from changes of the observed rotation period with the phase of the magnetic activity cycle. Stellar butterfly diagrams can be interpreted as the consequence of horizontal shear and a variation of the active latitude over the cycle, as observed in case of the Sun. Recently, differential rotation has been detected spectroscopically for a number of rapidly-rotating F stars by Reiners \& Schmitt (2003a, 2003b). Finally, surface spot rotation has been observed for a number of stars by Doppler imaging (Collier Cameron 2002; Kov\'ari et al.~2004).

One of the key questions in understanding both surface rotation and magnetic activity is what determines the rotation pattern, the surface shear, and the meridional flow pattern. Observational studies have so far focused on finding a relation between surface shear and the mean rotation period. The results found suggest a rather weak dependence of the type
\begin{equation}
 \delta \Omega \propto \Omega_0^n,
\end{equation}
with the values for $n$ derived by photometry and spectroscopy ranging from 0.3 to 0.8 (Henry et al.~1995; Donahue et al.~1996; Messina \& Guinan 2003) while Doppler imaging does not indicate any dependence on the rotation rate (Barnes et al.~2004).

The problem of stellar differential rotation is closely linked with that of the meridional flow. On the one hand the shear along the axis of rotation is one of the forces driving the flow, on the other hand the flow is a powerful transporter of angular momentum. Both the flow and the shear are key ingredients in the current theory of the solar dynamo (Durney 1995, Choudhuri et al.~1995, K\"uker et al.~2001). The shear in the overshoot layer beneath the convection zone generates of the toroidal field whereas the meridional flow at the bottom of the CZ produces a horizontal drift of the field belts towards the equator and thus the butterfly diagram. For this dynamo to work a counter-clockwise circulation with an amplitude of the order 10 m/s is needed. 

The meridional flow in the surface layer of the solar convection is directed towards the poles, with an amplitude of 20 m/s (Zhao \& Kosovichev 2004). For depths greater than 12 Mm and stars other than the Sun the flow is unknown. Observations of stellar activity cycles show solar-type as well as anti-solar butterfly diagrams. The latter can be interpreted in two ways, either as a solar-type dynamo with the active latitudes moving towards the equator or a different type of dynamo with the activity belts moving towards the poles. The first case would imply anti-solar rotation, i.e.~a shorter rotation period at the poles than at the equator. Solar-type rotation, on the other hand, would imply a different type of dynamo.
\section{The model}
Recent models by Kitchatinov \& R\"udiger (1999) and of K\"uker \& Stix (2001) successfully reproduce the basic features of the observed rotation pattern. The main generator of rotational shear is the Reynolds stress, which in a rotating, stratified fluid generally has a non-diffusive component in the azimuthal direction. Detailed expressions for the stress tensor as a function of the mean gas motion have been derived by Kitchatinov \& R\"udiger (1993) and  Kitchatinov et al.~(1994). We use these expressions to solve the equation of motion for axisymmetric flow in an anelastic fluid. 

We use the mean field approach, where the velocity field is split into a mean and a fluctuating part:
\begin{equation}
  \vec{u}=\vec{\bar{u}}+\vec{u}'.
\end{equation}
The equations of motion for the mean gas motion is the Reynolds equation,
\begin{equation} \label{reynolds}
  \rho \left [ \frac{\partial \vec{\bar{u}}}{\partial t}
      + (\vec{\bar{u}}\cdot \nabla)
       \vec{\bar{u}} \right ] =  \\
           - \nabla \cdot (\rho Q)
           - \nabla \bar{p} + \rho \vec{g} + \nabla \cdot \pi
\end{equation}
where 
\begin{equation}
T_{ij} = - \rho Q_{ij} = - \rho \langle u_i' u_j' \rangle,
\end{equation} 
is the Reynolds stress. A similar ansatz for the temperature leads to the heat transport equation,
\begin{equation} \label{conv_transp}
     \nabla \cdot  (\vec{F}^{\rm conv} + \vec{F}^{\rm rad}) - \rho
c_p \vec{u} \cdot \vec{\beta} = 0,
\end{equation}
where the convective heat flux is given by
\begin{equation} \label{heat_flux}
 F_{i}^{\rm conv} = \rho c_p \langle u'_i T' \rangle.
\end{equation}
Convection is driven by the superadiabatic temperature gradient,
\begin{equation}
 \vec{\beta} = \frac{\vec{g}}{c_p} - \nabla \bar{T}.
\end{equation}
The importance of the global rotation for the convective transport is measured by the Coriolis number, $\Omega^*=2 \tau_c \Omega$, where $\tau_c$ is the convective turnover time.

Equations (\ref{reynolds}) and (\ref{conv_transp}) are solved for axisymmetric solutions using a finite-difference code described in K\"uker \& Stix (2001). The density is prescribed but not constant, i.e. the anelastic approximation 
\begin{equation}
\nabla \cdot (\rho \vec{\bar{u}})=0
\end{equation}
 holds. As boundary conditions we require that both boundaries be stress-free and prescribe the heat flux. The latter requirement fixes the temperature gradient on the boundaries, but not the value of the temperature itself.

The inclusion of the heat transport equation solves a problem known as the "Taylor number puzzle". In the solar convection zone the Taylor number,
\begin{equation}
  {\rm Ta} = \frac{4 \Omega^2 R^4}{\nu^2},
\end{equation}
where $\Omega$ is the rotation frequency, $R$ the solar radius, and $\nu$ the turbulence viscosity, is a very large number. With $\Omega=2.7\times 10^{-6}s^{-1}$, $R=7\times 10^{10} {\rm cm}$, and $\nu\approx 10^{12} {\rm cm/s}$, a value of $7 \times 10^8$ follows. In an isothermal fluid with a Taylor number this large the rotation rate would have to be constant on cylindrical surfaces aligned with the axis of rotation, as required by the Taylor-Proudman theorem. This is avoided by taking into account the anisotropy of the convective heat transport caused by the Coriolis force, which gives rise to a small horizontal temperature gradient and thus a baroclinic term in the equation of motion. 

The appearance of the baroclinic term changes the asymptotic form of the equation for the meridional flow for fast rotation from
\begin{equation}
 \frac{\partial \Omega}{\partial z} \approx 0
\end{equation}
to
\begin{equation}
 r \sin \theta \frac{\partial \Omega^2}{\partial z}-\frac{g}{r c_p} \frac{\partial s}{\partial \theta} \approx 0.
\end{equation}
Solar-type differential rotation can thus exist at much larger Taylor numbers than in the purely hydrodynamic case.

\section{Results}
The model is applied to a series of main-sequence stars with 1.2, 1.0, 0.7, and 0.4 solar masses, respectively. Table \ref{table1} summaries their overall properties. For each star the rotation and flow are computed as function of the mean rotation period, defined by dividing the angular momentum of the stellar convection zone by its moment of inertia. For the Sun, this value equals the surface rotation period at about 30 deg latitude. The model has been gauged by the choice of the viscosity parameter, $c_\nu$, as defined by the expression for the turbulence viscosity,
\begin{equation}
  \nu_t=c_\nu l u_c,
\end{equation}  
where $l$ is the mixing length and $u_c$ the convection velocity. 

\begin{figure}
   \includegraphics[width=3.5cm]{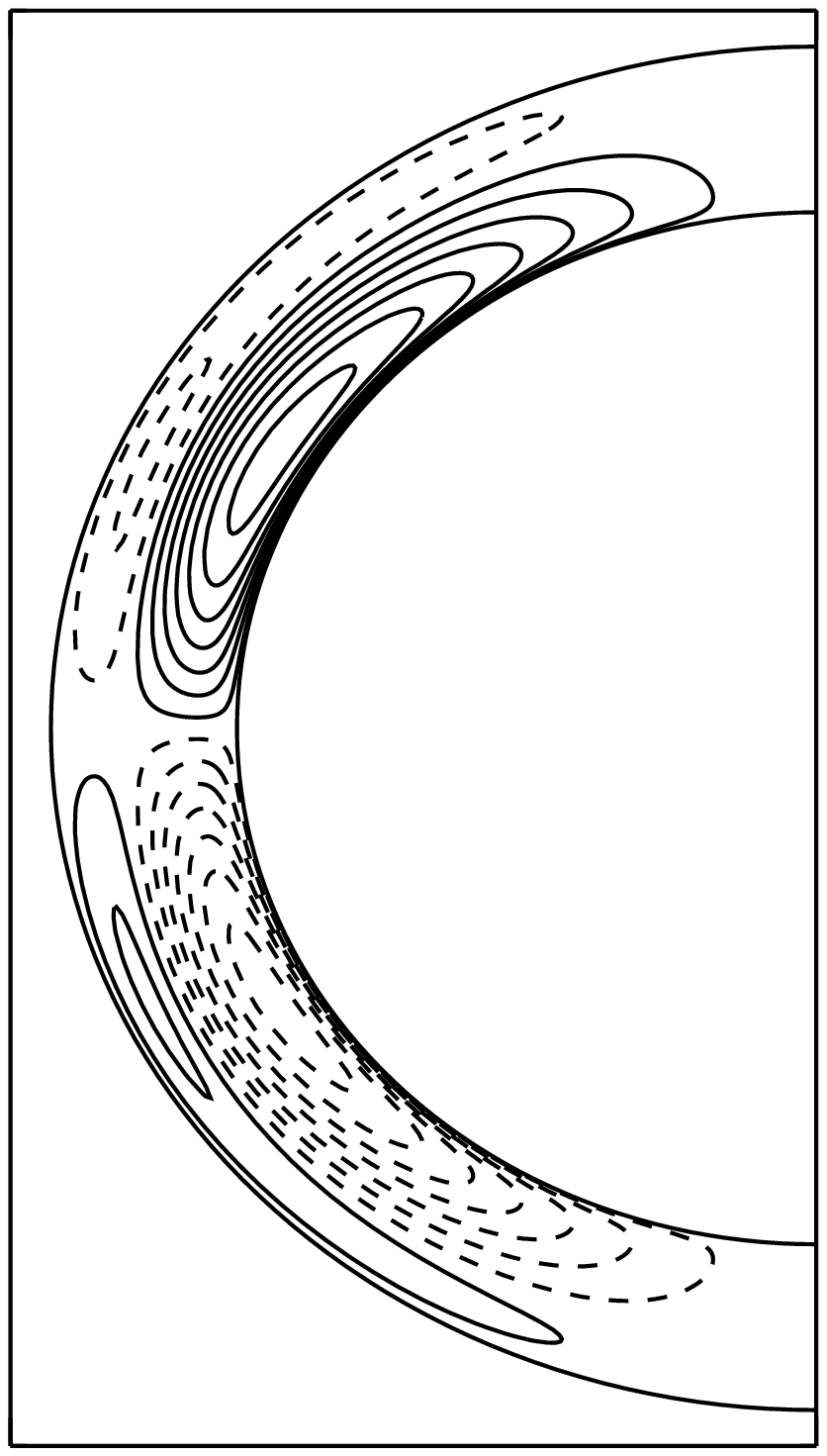} 
   \includegraphics[width=4.3cm]{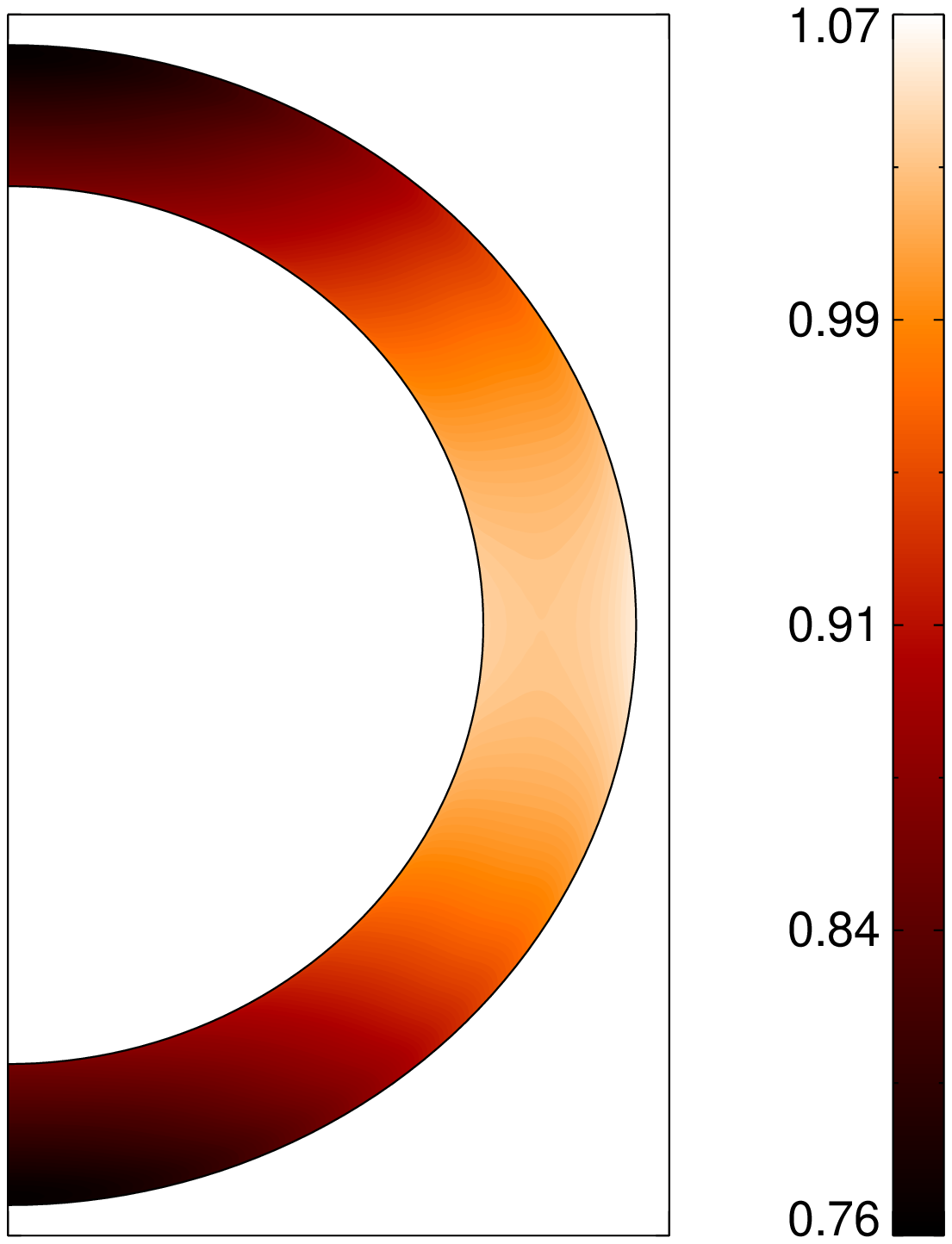} 
\caption{
    \label{sun}
  Solar meridional flow and DR. Mean rotation period is 27d.
}
\end{figure}

A series of computations with varying value of $c_\nu$ was carried out with a fixed rotation period of 27 days. A value of 0.15 yields the best reproduction of the solar surface shear and was therefore chosen for all subsequent computations. Figure \ref{sun} shows the (normalised) rotation rate (right diagram) and the meridional flow pattern resulting from this model for the Sun. The equator rotates about 30 percent faster than the poles, the variation with radius is weak, and the meridional flow shows two cells per hemisphere. A shallow cell of counter-clockwise flow (in the representation shown) resides in the upper part of the convection zone while a larger cell with clockwise flow occupies the remainder. The flow is towards the equator both at the bottom and the top of the convection zone and towards the poles at the interface between the two flow cells. The maximum flow speed is 4 m/s at the bottom and 6 m/s at the top of the convection zone.   

\begin{figure}
   \includegraphics[width=3.5cm]{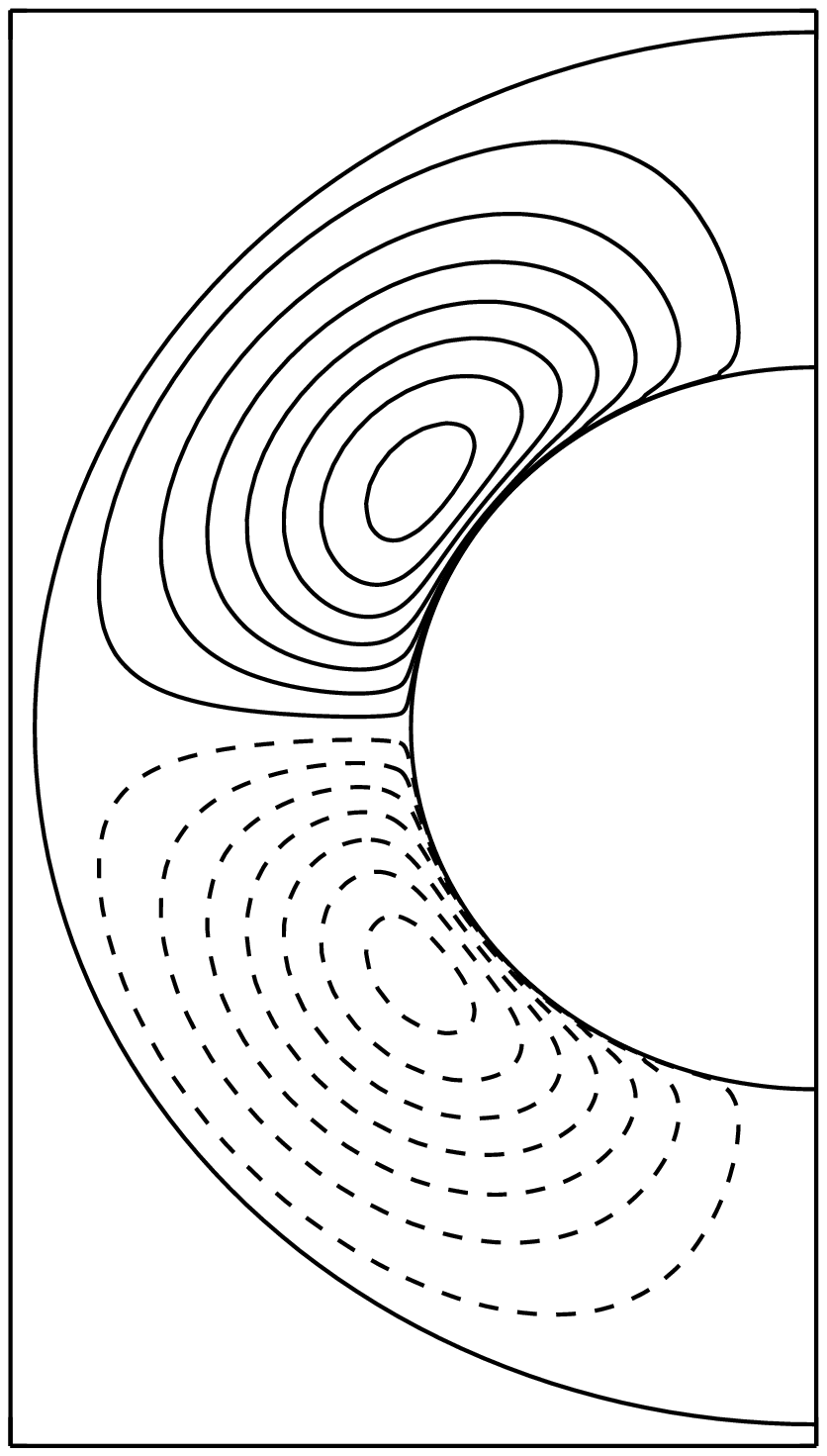} 
   \includegraphics[width=4.3cm]{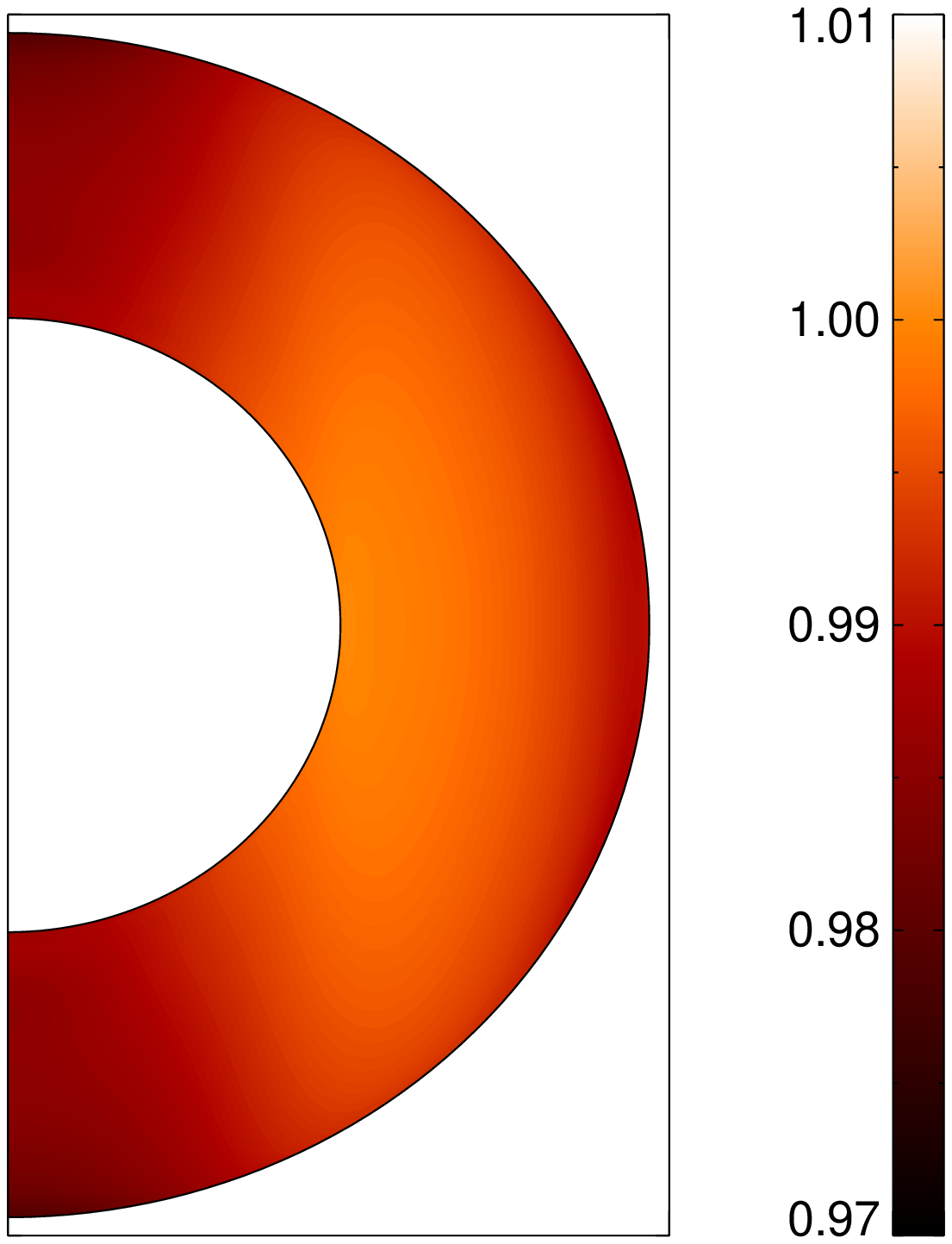} 
\caption{
    \label{mdwarf}
  Meridional flow and DR of an M dwarf of 0.4 solar masses. The mean rotation period is 10d.
}
\end{figure}

Figure \ref{mdwarf} shows the result for the 0.4 $M_\odot$ M dwarf rotating with a period of 10 days. The relative depth of the convection zone is greater in this type of star than for solar-type stars. The rotation is much more rigid than that of the Sun, showing only deviations up to two percent from the average rotation rate. The pattern is much more cylindrical than the solar rotation pattern, with slow rotation at the polar caps and (relatively) fast rotation in the cylinder surrounding the inner core. The meridional flow has only one cell per hemisphere, with the flow directed towards the equator at the bottom and towards the poles at the top of the convections zone, respectively.  The flow amplitude is 1.4 m/s at the bottom and 0.5 m/s at the top of the convection zone.

\begin{figure}
 \mbox{
 \hspace{-6mm}
 \includegraphics[width=4.5cm,height=5.0cm]{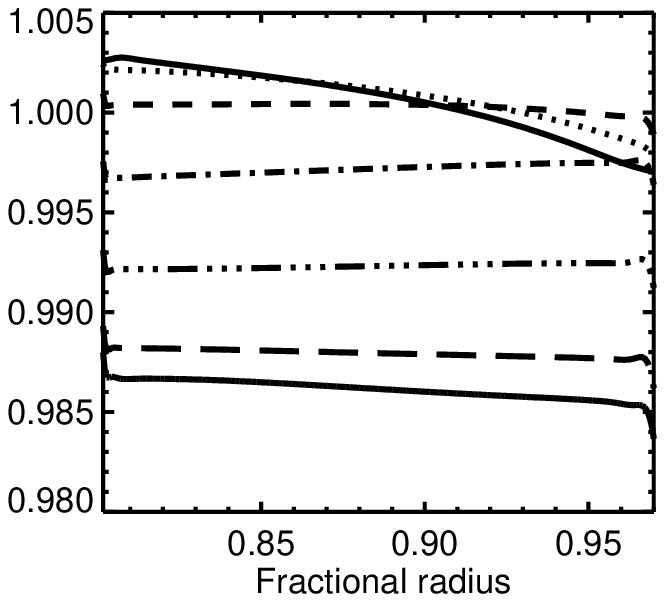}
 \includegraphics[width=4.5cm,height=5.0cm]{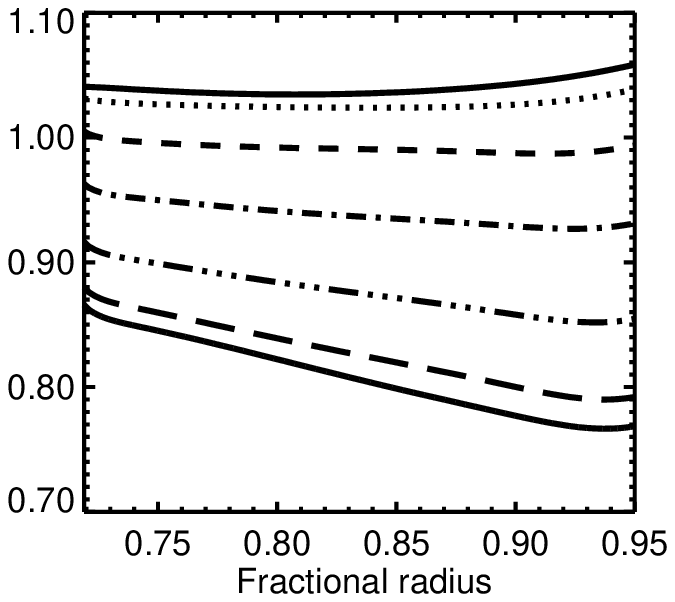}
 } 
  \mbox{ 
 \hspace{-6mm}
 \includegraphics[width=4.5cm,height=5.0cm]{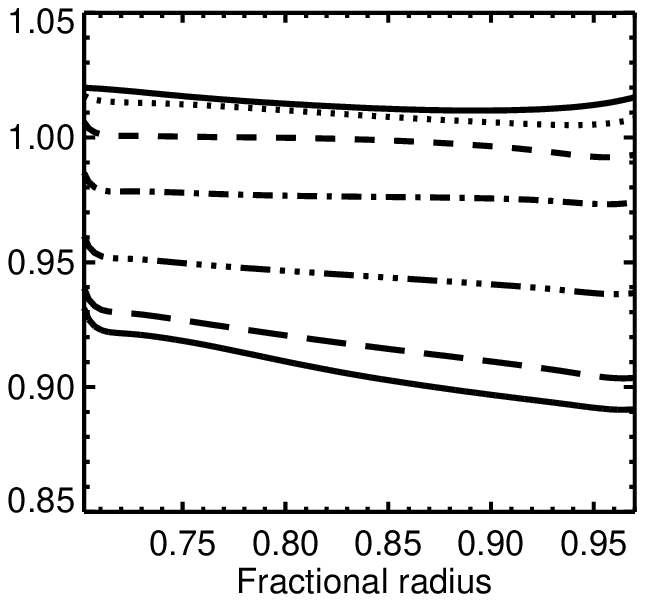}
 \includegraphics[width=4.5cm,height=5.0cm]{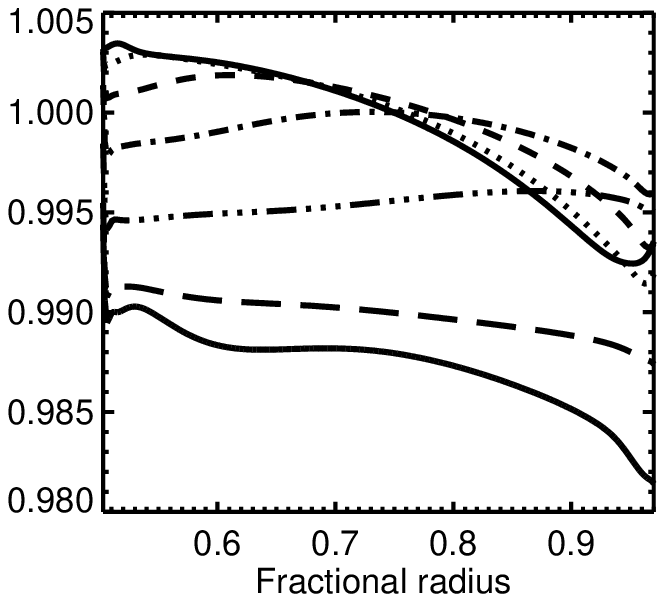}
 }
 \caption{ \label{grot}
   The normalised rotation period as a function of the fractional radius at the equator and at 15, 30, 45, 60, 70, and 90 degree latitude, respectively, from to top bottom in each diagram. Top left diagram: F star, rotating with a period of 1d. Top right: solar-type star, rotation period = 27d. Bottom left: K dwarf with a rotation period of 17 d. Bottom right: M dwarf, rotation period = 10d.
}
\end{figure}

Figure \ref{grot} shows the rotation rate vs.~radius for the four types of star at different latitudes. The relative shear is greatest for the solar-type star. As it rotates much faster, however, the F star has the largest value of the absolute surface shear. In all four cases the rotation is solar-type, i.e. the equator rotates faster than the poles. In all cases the radial shear is weak. The meridional flow at the bottom has an amplitude of 17.4 m/s for the F star and 2.8 m/s for the K dwarf. The corresponding values at the top are 33.2 m/s and 0.8 m/s. The surface temperature differences between poles and equator are 47.8 K for the F star, 1.7 K for the solar-type star, 0.4 K for the K dwarf, and 0.1 K for the K dwarf. In all cases the poles are hotter than the equator.

\begin{figure} \hspace{-0.5cm}
  \includegraphics[width=3.5cm]{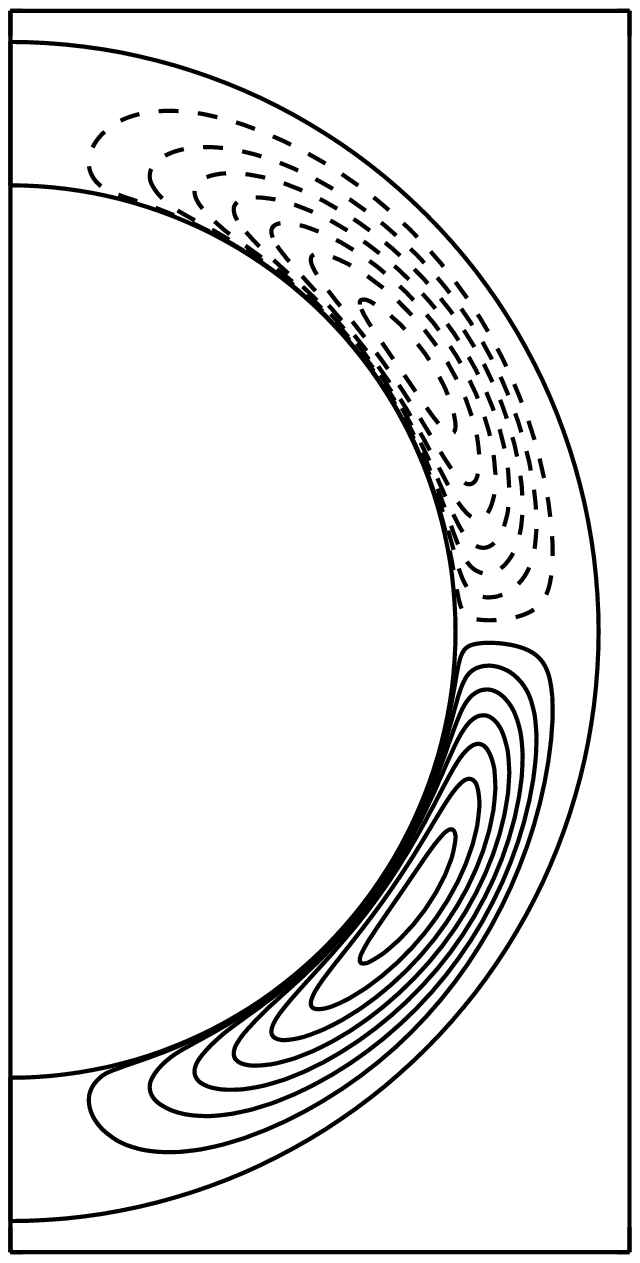} \hspace{-1cm}
  \includegraphics[width=3.5cm]{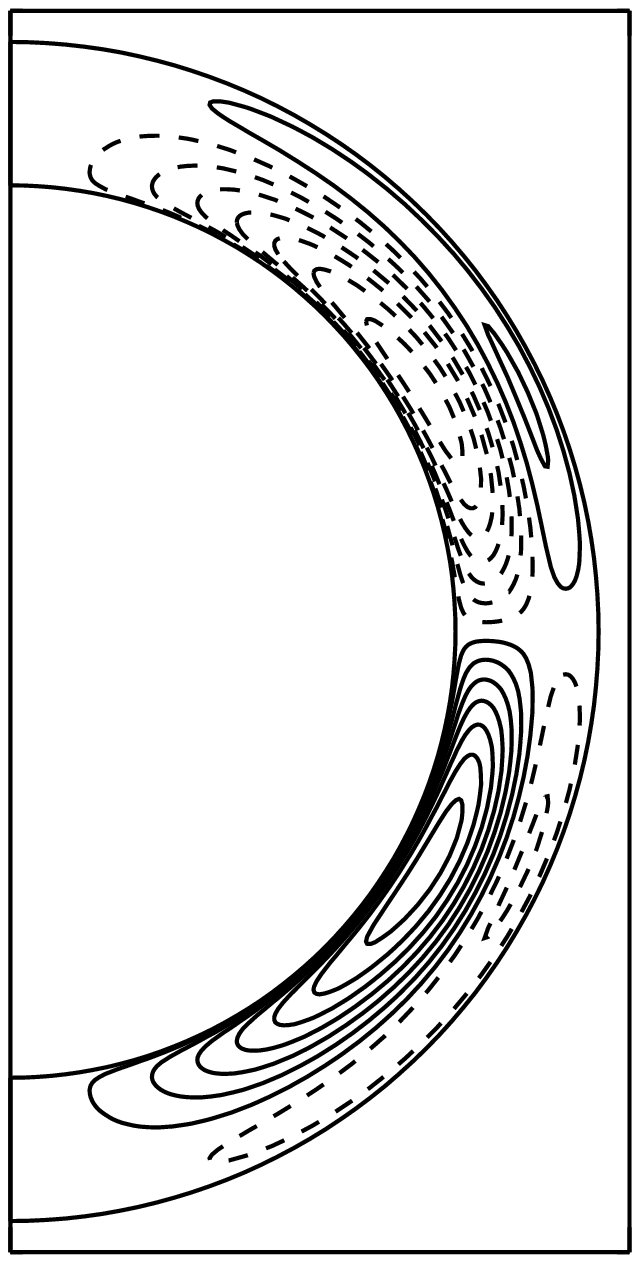} \hspace{-1cm}
  \includegraphics[width=3.5cm]{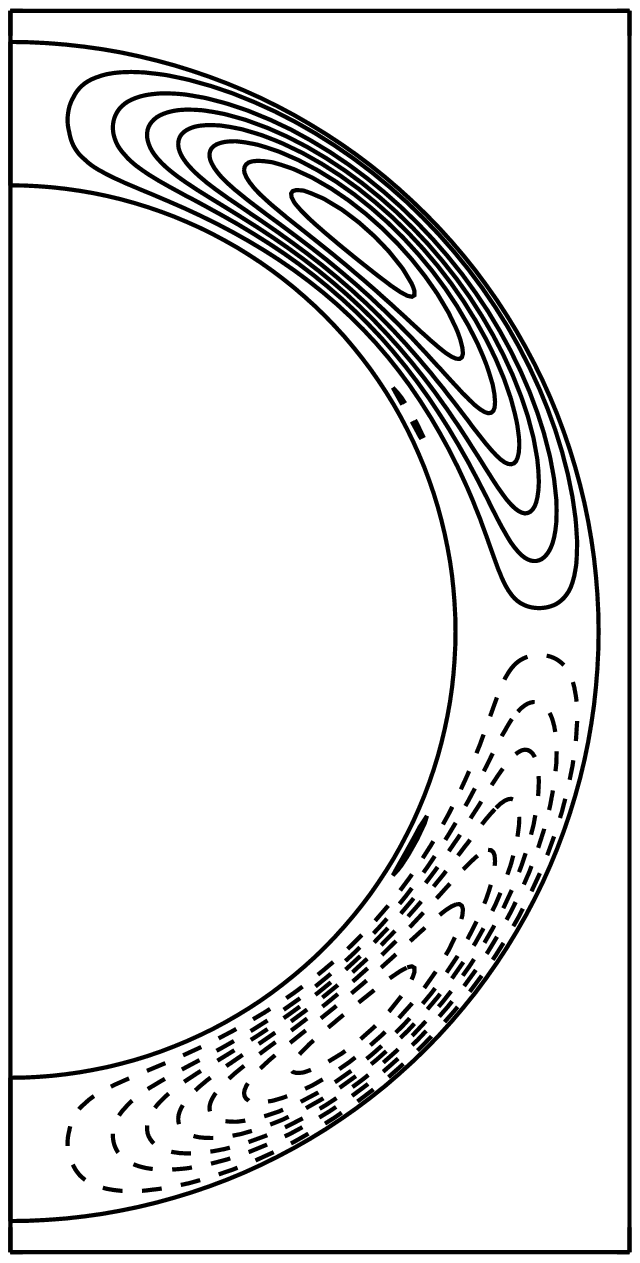}
  \caption{ \label{sunflowx}
   The meridional flow in the solar convection zone as resulting from our model. Solid lines denote  clockwise circulation, dashed lines counter-clockwise flow.
  }
\end{figure}

Figure \ref{sunflowx} shows the solar meridional circulation for three different rotation periods representing fast, intermediate, and slow rotation, respectively. The observed rotation period of 27 days is a case of intermediate rotation. The two-cell pattern found for this period is the result of the two driving forces, the shear along the z axis and the horizontal temperature gradient, being equally strong and opposing each other. The result is a clockwise flow cell in the upper part (where the Coriolis number is small and the rotation thus "slow", and a cell of counter-clockwise flow in the lower part, where the Coriolis number is large and the rotation thus "fast".

\begin{figure}
   \includegraphics[width=8.0cm]{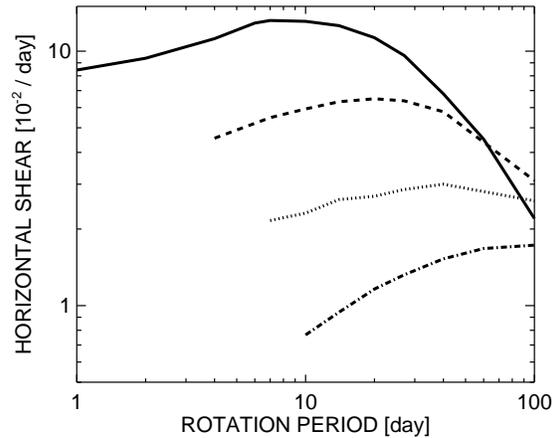} 
 \caption{
      \label{dom}   
      Surface shear as a function of rotation period for several types of main sequence star. Solid line: F star, dashed: solar-type, dotted: K dwarf, dash-dotted: M dwarf
 }
\end{figure}

Figure \ref{dom} summaries our findings concerning the surface shear. The total shear turns out to be a function of both rotation rate and spectral type, with the latter dependence being stronger. Each type of star has a maximum shear at certain rotation period, around which there is little variation.  
%
%
%
%
\begin{table}[h]
\caption{stellar properties}
\label{table1}
\begin{tabular}{ccccc}\hline 
Spectral Type & Mass &  Radius & $T_{\rm eff}$ & $\delta \Omega$\\
   & ($M_\odot$) & ($R_\odot$)& (K) & $10^{-2}$ s$^{-1}$\\
\hline 
M & 0.4 & 0.379 & 3518 & 1.7\\
K & 0.7 & 0.644 & 4347 & 3.0\\
G & 1.0 &  1.0  & 5777 & 6.4\\
F & 1.2 & 1.13  & 6236 & 13.0\\
\hline
\end{tabular}
\end{table}
\section{Discussion}
The surface meridional flow found for the Sun with its equatorward surface flow is in contradiction with the observations. The discrepancy is caused by the upper flow cell, caused by the small value of the Coriolis number in the top layer of the convection zone. The Kitchatinov \& R\"udiger (1993) expressions for the $\Lambda$ effect lead to positive radial shear in case of slow rotation, while the observed shear is negative. Box simulations by Chan (2001) and K\"apyl\"a et al.~(2004) found negative values for $\Lambda$, in agreement with the observations. We conclude that the Kitchatinov \& R\"udiger (1993) expressions are invalid for Coriolis numbers smaller than one. The flow at the bottom of the convection zone is in agreement with the requirements of the advection-dominated dynamo.

We find solar-type differential rotation in all cases. The total surface shear is mainly determined by the stellar luminosity and depends only weakly on the rotation rate. This confirms the observational finding by Barnes et al.~(2004)~who found a dependence of the type 
\begin{equation}
  \delta{\Omega}\propto T^{8.92\pm0.31}.
\end{equation}
The results from our model are summarised in Tab.~\ref{table1}. The values listed for $\delta \Omega$ are the maximum values for each star. Linear regression produces the power law 
\begin{equation}
  \delta{\Omega}\propto T^{3.28 \pm 0.52}.
\end{equation}
The exponent of 3.28 is considerably smaller than the value found by Barnes et al.~(2004) but  confirms the general trend. The dependence on the rotation rate shown in Fig.~\ref{dom} is weak. Because of the maximum, it does not follow a simple power law over the entire range of rotation periods for any of the stars. For the F, G, and K stars there is no significant dependence at all in the interval studied. Only the M dwarf shows a distinct increase of $\delta \Omega$ with the rotation period, i.e., the shear decreases with increasing rotation rate. This behaviour is well-known for purely hydrodynamic models where the meridional flow is driven by the shear alone and always tends to flatten the rotation profiles.  As Fig.~\ref{mdwarf} shows, the M dwarf is in the Taylor-Proudman state for a rotation periods of 10 days. Its small luminosity of 0.02 $L_\odot$ results in small convection velocities and hence large values of the Coriolis number ($\Omega^*=20$ in the bulk of the convection zone for $P$=10\,d) and of the Taylor number ($\sim 10^{12}$). Under these extreme conditions the stabilising effect of the horizontal temperature gradient is no longer sufficient to keep up the horizontal shear, and we find essentially rigid rotation, like in the pure hydrodynamic case (K\"uker \& R\"udiger 1997; R\"udiger et al.~1998)
%

\end{document}